\documentclass[12pt,epsf]{article}
\usepackage{amsmath}
\usepackage{amssymb}
\begin{document}
\def\a{\alpha}
\def\b{\beta}
\def\c{\varepsilon}
\def\d{\delta}
\def\e{\epsilon}
\def\f{\phi}
\def\g{\gamma}
\def\h{\theta}
\def\k{\kappa}
\def\l{\lambda}
\def\m{\mu}
\def\n{\nu}
\def\p{\psi}
\def\q{\partial}
\def\r{\rho}
\def\s{\sigma}
\def\t{\tau}
\def\u{\upsilon}
\def\v{\varphi}
\def\w{\omega}
\def\x{\xi}
\def\y{\eta}
\def\z{\zeta}
\def\D{{\mit \Delta}}
\def\G{\Gamma}
\def\H{\Theta}
\def\L{\Lambda}
\def\F{\Phi}
\def\P{\Psi}

\def\S{\Sigma}

\def\o{\over}
\def\beq{\begin{eqnarray}}
\def\eeq{\end{eqnarray}}
\newcommand{\gsim}{ \mathop{}_{\textstyle \sim}^{\textstyle >} }
\newcommand{\lsim}{ \mathop{}_{\textstyle \sim}^{\textstyle <} }
\newcommand{\vev}[1]{ \left\langle {#1} \right\rangle }
\newcommand{\bra}[1]{ \langle {#1} | }
\newcommand{\ket}[1]{ | {#1} \rangle }
\newcommand{\EV}{ {\rm eV} }
\newcommand{\KEV}{ {\rm keV} }
\newcommand{\MEV}{ {\rm MeV} }
\newcommand{\GEV}{ {\rm GeV} }
\newcommand{\TEV}{ {\rm TeV} }
\def\slash#1{\ooalign{\hfil/\hfil\crcr$#1$}}
\def\diag{\mathop{\rm diag}\nolimits}
\def\Spin{\mathop{\rm Spin}}
\def\SO{\mathop{\rm SO}}
\def\O{\mathop{\rm O}}
\def\SU{\mathop{\rm SU}}
\def\U{\mathop{\rm U}}
\def\Sp{\mathop{\rm Sp}}
\def\SL{\mathop{\rm SL}}
\def\tr{\mathop{\rm tr}}

\baselineskip 0.7cm

\begin{titlepage}

\begin{flushright}
UCB-PTH-08/61
\end{flushright}

\vskip 1.35cm
\begin{center}
{\large \bf An Index for Non-relativistic Superconformal Field Theories}
\vskip 2.2cm

Yu Nakayama

\vskip 0.4cm
{\it Berkeley Center for Theoretical Physics and Department of Physics}
\\

\vskip 3.5cm

\abstract{We study the highest-weight representation of $\mathcal{N}=2$ supersymmetric Schr\"odinger algebra which appears in non-relativistic superconformal field theories in $(1+2)$ dimension. We define the index for the non-relativistic superconformal field theories and study its properties. As a concrete example, we compute the index for the non-relativistic limit of  $\mathcal{N}=6$ superconformal Chern-Simons-matter theory recently proposed by Aharony et al.}
\end{center}
\end{titlepage}

\setcounter{page}{2}

\section{Introduction}
It is well-known but rather mysterious that Schr\"odinger equation for a free particle  has a hidden dynamical symmetry in addition to the obvious Galilean invariance of a classical particle \cite{Hagen:1972pd}\cite{Niederer:1972zz}\cite{Perroud:1977qh}\cite{Barut:1981mt}. It has a symmetry of conformal invariance (dilatation and special conformal transformation) and some interesting properties of this enhanced symmetry have been studied in the literatures \cite{Hussin:1986cc}\cite{Jackiw:1990mb}\cite{Jackiw:1992fg}\cite{Henkel:1993sg}\cite{Dobrev}\cite{Mehen:1999nd}\cite{Nishida:2007pj}. Furthermore, this non-relativistic conformal symmetry can be realized in non-relativistic conformal field theories, and they may play an important role in condensed matter physics such as the quantum Hall effect and high $T_c$ superconductivity.

It is not a totally absurd idea (beyond theoretical curiosity) to investigate the supersymmetric extension of the non-relativistic conformal algebra. First of all, in our real nature, the supersymmetry was first discovered not in the  frontier of high energy physics, but in the realm of condensed matter physics. For example, the tri-critical Ising model at the critical point possesses a dynamical (or accidental) superconformal symmetry and can be realized in adsorbed helium on krypton-plated
graphite \cite{exp}. It is quite plausible that a similar construction realizes non-relativistic superconformal symmetry.

The representation theory and the invariant quantities should play fundamental roles in classifying such non-relativistic superconformal field theories. 
In this paper, we study the highest-weight representation of the non-relativistic superconformal algebra in $(1+2)$ dimension. The unitary (projective) representation of the non-relativistic conformal algebra has been studied in the literature \cite{Perroud:1977qh}, but our prescription (see also \cite{Dobrev} for $(1+1)$ dimensional case) turns out to be more useful when applied to operator-state correspondence in the non-relativistic superconformal field theories \cite{Nishida:2007pj}. As a concrete application, we define the superconformal index, which is invariant under any marginal deformation of the non-relativistic superconformal field theories. 

The index defined as a super-trace (i.e. graded trace with respect to the fermion number) over the Hilbert space is a very robust quantity \cite{Witten:1982df}. In any proposed duality, it would be the first check to compare the index computed from both sides. Furthermore, the index itself often possesses interesting mathematical properties such as modular properties or relation to the underlying geometry. As in the relativistic case, our superconformal index will be a unique candidate for the invariants of the non-relativistic superconformal field theories.

Recently, a new class of relativistic superconformal field theories in $(1+2)$ dimension with $\mathcal{N}=6$ supersymmetry has been proposed by Aharony et al \cite{Aharony:2008ug} (ABJM model). As a concrete example of our superconformal index, we compute the index for the non-relativistic limit of such theories in the large $N$ 't Hooft limit.

The organization of the paper is as follows. In section 2, we review $\mathcal{N}=2$ non-relativistic superconformal algebra in $(1+2)$ dimension. In section 3, we investigate the highest-weight representation of the non-relativistic superconformal algebra. In section 4, we introduce the concept of the index and study its basic properties. In section 5, we present explicit computation of the index for some non-trivial $(1+2)$ dimensional non-relativistic superconformal field theories. In section 6, we give some discussions for future directions.

\section{Non-relativistic Superconformal Algebra}
$\mathcal{N}=2$ superconformal algebra in $(1+2)$ dimension has been first studied in \cite{Leblanc:1992wu} and recently re-derived by a reduction from $SU(2,2|1)$ relativistic conformal algebra in \cite{Sakaguchi:2008ku} (see also \cite{Gauntlett:1990xq}\cite{Horvathy:1993fd}\cite{Duval:1993hs}). The commutation relation is summarized in the following form. For a bosonic part, we have (we will omit trivial commutation relations)
\begin{align}
i[J,P] &= -i P \ , \ \ i[J,\bar{P}] = i \bar{P} \ , \ \ i[J,G] = - iG \ , \ \ i[J,\bar{G}] = i\bar{G} \ ,\cr
i[H,G] &= P \ , \ \ i[H,\bar{G}] = \bar{P} \ , \ \ i[K,P] = - G \ ,  \ \ i[K,\bar{P}] = - \bar{G} \ ,\cr
i[D,P] &= - P \ , \ \ i[D,\bar{P}] = - \bar{P} \ , \ \ i[D,G] = G \ , \ \ i[D,\bar{G}] = \bar{G} \ ,\cr
 i[D,H] &= -2H \ , \ \ i[H,K] = D \ , \ \ i[D,K] = 2K \ , \ \ i[P,\bar{G}] = 2M \ , 
\end{align}
where $J = J_{12}$ is $U(1)$ angular momentum, $P = P_1+iP_2$ and $\bar{P} = P_1 - iP_2$ are spatial momenta, $H$ is the Hamiltonian, $G = G_1+iG_2$ and $\bar{G} = G_1-iG_2$ are Galilean boost, $D$ is the dilatation, $K$ is the special conformal transformation, and $M$ is the mass operator which is the center of the non-relativistic conformal algebra. 

The simple way to obtain this bosonic part of the non-relativistic conformal algebra (also known as Schr\"odinger algebra) is to embed it in one-dimensional higher relativistic conformal algebra (in our case $SO(2,4)$). Note that it does not have a simple relation to the relativistic conformal algebra with the same space-time dimension because relativistic conformal field theories where everything travels at the speed of light do not have a definite non-relativistic limit. Probably the most mysterious special conformal transformation of the Schr\"odinger algebra is generated by $\delta t = - \epsilon t^2$ and $\delta r_i = - \epsilon tr_i$. This corresponds to the special conformal transformation of $SO(2,4)$ in a particular light cone direction.

All these generators are realized in a Hermitian basis in the sense that there is an involutive anti-automorphism of the algebra generated by\footnote{$w_0$ by the dagger here means the conventional hermitian conjugation used in quantum mechanics.} $w_0(J)= J^\dagger = J$, $w_0(P)=P^\dagger = \bar{P}$, $w_0(G)=G^\dagger = \bar{G}$, $w_0(H) = H^\dagger = H$, $w_0(D) = D^\dagger = D$, $w_0(K) = K^\dagger = K$, and $w_0(M) = M^\dagger = M$. We will encounter a different anti-automorphism later. These anti-automorphisms can be used to define the conjugate states and hense to define the inner product; here we have two different ways to define them in accordance with physical applications as we will see.

$\mathcal{N}=2$ superconformal algebra is augmented by three (complex) fermionic generators $(Q_1,Q_1^*, Q_2,Q_2^*,S,S^*)$ with bosonic R-symmetry generator $R$ as
\begin{align}
\{ Q_1, Q_1^* \} &= 2M \ , \ \ \{Q_2,Q_2^*\} = H \ , \ \ \{Q_1,Q_2^*\} = \bar{P} \ , \ \ \{Q_2,{Q}_1^*\} = P \ ,\cr 
i[J,Q_1] &= \frac{i}{2}Q_1 \ , \ \ i[J,Q_1^*] = -\frac{i}{2}Q_1^* \ , \ \ i[J,Q_2] = -\frac{i}{2}Q_2 \ , \ \ i[J,Q^*_2] = \frac{i}{2} Q_2^* \ , \cr
i[\bar{G},Q_2] &= -Q_1 \ , \ \ i[G,Q_2^*] = -Q_1^* \ , \ \ i[D,Q_2] = -Q_2 \ , \ \ i[D,Q_2^*] = -Q_2^* \ , \cr
i[K,Q_2] &= S \ , \ \ i[H,S^*] = -Q_2^* \ , \ \ i[\bar{P},S] = -Q_1 \ , \ \ i [J, S] = -\frac{i}{2} S \ , \cr
\{S,S^*\} &= K \ , \ \ \{S,Q_1^*\} = -G \ , \ \ i[D,S] = S \ , \ \ \{S,Q_2^*\} = \frac{i}{2}(iD - J +\frac{3}{2}R) \ , \cr
i[R,Q_a] &=  -i Q_a \ , \ \ i[R,S] = -iS \ ,
\end{align}
where $w_0(Q_1)=Q_1^*$, $w_0(Q_2) = Q_2^*$ and $w_0(S) = S^*$. 

We note that the non-relativistic superconformal algebra $\mathrm{SSch}_2$ has a grading structure with respect to the dilatation $D$ and can be triangular-decomposed as $\mathrm{SSch}_2 = S_+ \oplus S_0 \oplus S_-$, where $S_+ = \{ P ,\bar{P},H, Q_2, Q_2^*\}, \  S_0 = \{ J, R, M, D, Q_1, Q_1^*\} , \ S_- = \{G, \bar{G}, K, S, S^*\}$. Relatedly, for later purposes, we notice that $\mathrm{SSch}_2$ has a non-trivial involutive anti-automorphism of the algebra given by $w(J)= J$, $w(P)= \bar{G}$, $w(G)=\bar{P}$, $w(H) =-K$, $w(K) = -H$, $w(D) = -D$,  $w(M) = -M$, $w(R) = R$, $w(Q_1) = iQ_1^*$, $w(Q_1^*)=iQ_1$, $w(Q_2) = iS^*$, $w(Q_2^*) = iS$, $w(S) = iQ_2^*$ and $w(S^*) = iQ_2$. This anti-automorphism is much like Belavin-Polyakov-Zamolodchikov (BPZ) conjugation of the relativistic conformal field theories \cite{Belavin:1984vu} and plays a central role in the representation theory of the non-relativistic superconformal algebra as we will see.\footnote{Mathematically, the corresponding BPZ inner-product is known as the Shapovalov form \cite{sha}, which is useful to analyse the reducibility of the representation.} Physically, this is related to the Euclidean version of the conformal algebra and state-operator correspondence (see e.g. \cite{Nishida:2007pj} for some discussions).

\section{Highest Weight Representation}
In this section, we study the highest-weight representation of $\mathrm{SSch}_2$.
\subsection{Without SUSY}
We begin with the highest-weight representation of the bosonic Schr\"odinger algebra $\mathrm{Sch}_2$.\footnote{(1+1) dimensional case $\mathrm{Sch}_1$ was discussed in \cite{Dobrev}.} The grading structure mentioned in section 2 suggests that it is convenient to choose the Cartan subalgebra $D,J, M$ and diagonalize them. We denote the corresponding eigen-state as $|d,j,m\rangle$, where
\begin{eqnarray}
iD | d, j,m\rangle = -d| d, j,m\rangle \ ,\ \ J | d, j,m\rangle = j | d, j,m\rangle \ , \ \ iM | d, j,m\rangle  = m | d, j , m \rangle \ .
\end{eqnarray}
It might seem peculiar at first sight that the  ``Hermitian" generator $D$ (and also $M$) has a pure imaginary eigenvalue, but this happens because of the lack of the finite norm defined by $w$ instead of $w_0$. More physically, it is related to the Wick rotation of the Schr\"odinger algebra in the radial quantization scheme and most easily seen by the structure of the  anti-automorphism $w(D) = -D$ compatible with the state-operator correspondence.\footnote{The precisely same situation occurs in relativistic conformal algebra: see e.g. \cite{Minwalla:1997ka}\cite{Dolan:2002zh}.}

By definition, the highest-weight state (called primary states in \cite{Nishida:2007pj} and quasi-primary in \cite{Henkel:1993sg}) is a state annihilated by all the raising operators $G,\bar{G}$ and $K$. Descendant states or simply descendants are created by acting $P,\bar{P}$ and $H$ on the highest-weight state. More explicitly, descendants are spanned by the states
\begin{eqnarray}
P^x \bar{P}^y H^z |d_0,j_0, m \rangle \ ,
\end{eqnarray}
which have quantum numbers $d = d_0 + x + y + 2z$ and $j = j_0 - x + y$. 

Not all descendants are physical states: in order to construct irreducible representations, we have to subtract null vectors in the Verma module. The singular vectors are given by the descendants which are also annihilated by $G,\bar{G}$ and $K$. With this regard, it is not so difficult to prove the following proposition:\footnote{We always assume $m\neq 0$ in this paper. The representation theory becomes slightly complicated for $m=0$.}

{\bf Proposition 1} 

The singular vectors of the Verma module over $\mathrm{Sch}_2$ are given by (up to a trivial overall factor)
\begin{eqnarray}
 \sum_{n=0}^k a_n P^n \bar{P}^n H^{k-n} |d_0,j_0,m\rangle 
\end{eqnarray}
with $-d_0 = k-2$. The coefficient $a_n$ is given by  
\begin{eqnarray}
a_n = \frac{(k-n)!}{n!} \frac{1}{(2mi)^n} \ . \label{gamma}
\end{eqnarray}

The proof is based on the direct computation. Without losing generality, we assume that from the charge conservation for $j$ and $d$, a singular vector is given by $\sum_{n=0}a_n P^{n+a}\bar{P}^n H^{k-n}$. Acting $K$ gives the condition
\begin{eqnarray}
a_{n+1} 2m (n+1)(n+1+a) + ia_n (k-n)(k+n-1+a+d_0) = 0\ .
\end{eqnarray}
On the other hand,
acting $\bar{G}$ gives the condition
\begin{eqnarray}
a_{n+1} 2m (n+1+a) + ia_n(k-n)  = 0
\end{eqnarray}
while acting $G$ gives the condition
\begin{eqnarray}
a_{n+1}2m (n+1) + ia_n (k-n)  = 0 \ .
\end{eqnarray}
The consistency of these three recursion relations demands $a=0$ and $-d_0= k-2$;  then the recursion relation is solved as in \eqref{gamma}.

For example, at $k=1$, we have a free Schr\"odinger equation (or Euclidean heat equation) $H|\Psi\rangle  =- \frac{P\bar{P}}{2mi}|\Psi\rangle $. Similarly we can construct hierarchy of higher order equations of motions compatible with the Schr\"odinger algebra by studying the singular vectors at different $k = -d_0+2$. Note that the singular vector appears only for specific quantized values of $d_0 = 1, 0, -1, -2 ,\cdots$. 

One can also investigate the structure of the singular vectors by studying the determinant of the matrix of all Shapovalov forms at a fixed level $k$ and angular momentum $j$. We conjecture and indeed it can be checked by a direct computation at lower levels, it is proportional to $(d_0-1)^{k}d_0^{k-1} (d_0+1)^{k-2} \cdots (d_0-2+k)$. This is in accord with the singular vector structure discussed above. As a simple application, this determinant formula suggests the (necessary) unitarity condition $d\ge1$ from the positive definiteness of the Shapovalov forms.

\subsection{With SUSY}
The supersymmetric analogue of the highest-weight representation is given by the states annihilated by $G,\bar{G},K,S,S^*$ and $Q_1^*$. The descendants are obtained by acting $P,\bar{P}, H , Q_2,Q_2^*$ and $Q_1$ on highest-weight state. Since all the states are further classified by the highest-weight representation of the bosonic subalgebra, we can focus on the states spanned by $Q_1^\alpha Q_2^\beta Q_2^{*\gamma} |X\rangle$, where $\alpha, \beta, \gamma = 0,1$ from Fermi statistics. Because of the R-symmetry, the state has an extra conserved quantum number $r$. A general long multiplet has $8$ components, but other short multiplets are possible (in addition to the possibility of multiplet shortening in the bosonic descendants). By directly acting $G,\bar{G},K,S,S^*$ and $Q_1^*$ on $Q_1^\alpha Q_2^\beta Q_2^{*\gamma} |X\rangle$, we can easily classify all the irreducible representations of $\mathrm{SSch}_2$. We summarize the result in the following form:

{\bf Proposition 2}

There are four types of unitary representations of $SSch_2$.
\begin{enumerate}
	\item The vacuum representation is annihilated both by $Q_2$ and $Q_2^*$: it is given by $|d_0 = 0, j_0 = \frac{3}{2} r_0 , m \rangle$.
	\item Chiral representation is annihilated by $Q_2^*$: From the anti-commutation relation $\{S,Q_2^*\} = \frac{i}{2}(iD-J+\frac{3}{2}R)$, the states should satisfy $d_0 =  -j_0 + \frac{3}{2} r_0 $ . 
 	\item Anti-chiral representation is annihilated by $Q_2$: From the anti-commutation relation $\{S^*,Q_2\} = -\frac{i}{2}(-iD-J+\frac{3}{2}R)$, the states should satisfy $d_0 = j_0 - \frac{3}{2}r_0 $ .
	\item A long multiplet is annihilated neither by $Q_2$ nor $Q_2^*$.
\end{enumerate}
The branching rule is very simple, the two chiral representations (or two anti-chiral representations) form a long multiplet. It is also useful to note that acting ${P}$ ($\bar{P}$) does not affect the chiral (anti-chiral) condition.

It is important to note that the condition does not involve any quantization of the scaling dimension $d$ unlike the case for the bosonic short representation. In this sense, the structure is closer to the representation theory of the relativistic superconformal algebra \cite{Dobrev:1985qv}.

\section{Index for Non-relativistic Superconformal Field Theories}
A concept of the index is of great significance in the classification of supersymmetric theories because it is invariant under any small perturbation (in our case, exactly marginal deformation) of a given theory. The definition depends on the inner product of the Hilbert space we study and in particular it depends on the anti-automorphism of the algebra to define the conjugate state.

For our purposes, the simple choice would be to use the anti-automorphism defined by $w_0(O) = O^\dagger$ introduced in section 2. The corresponding index should be\footnote{We could introduce chemical potential whose charge commutes with $Q_2$.}
\begin{eqnarray}
I_0 = \mathrm{Tr} (-1)^F e^{-\beta H} \ ,
\end{eqnarray}
where $H= \{w_0(Q_2^*), Q_2^*\} = \{Q_2 , Q_2^*\}$. The trace is defined by the anti-automorphism given by $w_0$: $\mathrm{Tr}O = \sum_i \langle i|O|i\rangle = \sum_i \langle 0|w_0(O_i) O O_i|0\rangle$. 
Because of the Bose-Fermi cancellation, the index does not depend on $\beta$ and just counts the zero-energy state (``vacuum" in the usual sense). This is nothing but the Witten index of the supersymmetric quantum mechanics \cite{Witten:1982df}.

A more non-trivial choice is to use the non-trivial anti-automorphism defined by $w$ (BPZ conjugation) introduced in section 2. Then, the index has contributions from all the  short multiplets of the non-relativistic superconformal algebra studied in section 3.2. We now define
\begin{eqnarray}
I(x) = \mathrm{Tr} (-1)^F e^{-\beta \Delta} x^{R-2J} \ ,
\end{eqnarray}
where $\Delta = \{w(Q_2^*), Q_2^*\} = i\{S,Q_2^*\} = -\frac{1}{2}(iD-J+\frac{3}{2}R)$. Here the trace is defined by the anti-automorphism $w$ introduced in section 2: $\mathrm{Tr}O = \sum_i \langle i|O|i\rangle = \sum_i \langle 0|w(O_i) O O_i|0\rangle$. We have introduced the chemical potential $x$ to distinguish each contributions to the index from different (actually infinitely many in our later examples) short multiplets. Note that the corresponding generator $R-2J$ commute with $Q_2^*$. The unitarity assumption for the inner product (realized by the BPZ conjugation $w$) guarantees that $\Delta \ge 0$. A similar superconformal index in the relativistic field theories has been studied in the literatures (see e.g. \cite{Romelsberger:2005eg}\cite{Kinney:2005ej}\cite{Nakayama:2005mf}\cite{Nakayama:2006ur}\cite{Nakayama:2007jy}\cite{Romelsberger:2007ec}\cite{Dolan:2008qi}\cite{Bhattacharya:2008bja}).

The contribution to the index only comes from states with $\Delta =0$ or in other words, the states annihilated by $Q_2^*$: they are chiral primaries (and their descendants generated by ${P}$). As is clear from the branching rule discussed in section 3, this index does not depend on the exact marginal deformation of the theory because the contribution is zero when two chiral multiplets merge into a long multiplet (or when a long multiplet decomposes into two chiral multiplets). 

\section{Examples}
We present several examples of explicit computation of the index for non-relativistic superconformal field theories.
\subsection{$\mathcal{N}=2$ superconformal Abelian Chern-Simons-matter theory}
We first consider the $\mathcal{N}=2$ superconformal Abelian Chern-Simons-matter model in the non-relativistic limit \cite{Leblanc:1992wu}.
The action is given by
\begin{align}
S &= \int dt d^2x \left[ \frac{\kappa}{2}\partial_t A_i \epsilon_{ij} A_j - A_0 \left(\kappa B + e|\Phi|^2 + e|\Psi|^2\right)   + i\Phi^* \partial_t \Phi + i \Psi^*\partial_t \Psi \right. \cr 
&- \left. \frac{1}{2m}|D_i\Phi|^2 -\frac{1}{2m}|D_i\Psi|^2  + \frac{e}{2m}B|\Psi|^2 + \frac{e^2}{2m\kappa} |\Phi|^4 + \frac{3e^2}{2m\kappa}|\Phi|^2|\Psi|^2  \right]  \ ,
\end{align}
where $\Phi$ is a bosonic field and $\Psi$ is a fermionic field.

The index does not depend on the continuous parameter of the theory, so it would be convenient to take $\frac{e^2}{k}\to0$, where the theory is free. In the zero coupling limit,  We have explicit expressions for supersymmetry generators:
\begin{align}
Q_1 &= \frac{1}{\sqrt{2m}} \int d^2x \Phi^* \Psi \cr
Q_2 &= \frac{1}{\sqrt{2m}} \int d^2x \Phi^* (\partial_1 + i\partial_2) \Psi \ .  
\end{align}
It is easy to see $i[Q_1^*, \Phi] = i[Q_2^*, \Phi] = \{Q_1^*, \Psi^*\} = \{Q_2^*,\Psi^*\} = 0$. 

\begin{table}[tb]
\begin{center}
\begin{tabular}{c|c|c|c|c|c}
 Letters        & $U(1) $& $J$ & $R$ & $iD$ & $R-2J$ \\
 \hline
 $\Phi$&     $1 $         & $0$ & $2/3$ &$-1$ & 2/3 \\
  $\Psi^*$&     $-1 $         & $-1/2$ & $1/3$ &$-1$ & 4/3 \\
  $P$&     $0 $         & $-1$ & $0$ & $-1$ & 2 \\ 
 \end{tabular}
\end{center}
\caption{List of the letters contributing to the index (hence $\Delta = 0$) for $\mathcal{N}=2$ Chern-Simons-matter theory.}
\label{tab:1}
\end{table}%

Let us first forget the gauge invariance for a while and study the index for the non-gauge system (in the zero coupling limit). The propagating degrees of freedom is $\Phi$ and $\Psi$ (and their complex conjugate). It is easy to see that the chiral primary operators (states) are generated by $\Phi$ and $\Psi^*$ (see Table \ref{tab:1}). The chiral ring is spanned by $\mathcal{R} = \{ \Phi^m , \Psi^* \Phi^n \} $ for $(m,n=0,1,2,\cdots)$ and their derivatives by acting $\bar{P}$. The index is readily computed as
\begin{align}
I(x) = \mathrm{Tr}(-1)^F x^{R-2J} &= \exp\left(\sum_{n=1}^{\infty} \frac{1}{n} \frac{x^{\frac{2}{3}n}}{1-x^{2n}}- \frac{1}{n} \frac{x^{\frac{4}{3}n}}{1-x^{2n}}\right) \cr
&= \prod_{m=0}^{\infty} \frac{1-x^{\frac{4}{3}+2m}}{1-x^{\frac{2}{3}+2m}} \  .
\end{align}

In order to make contact with the non-zero coupling case, we should have restricted ourselves to gauge invariant states. The index can be computed as
\begin{align}
I(x) = \mathrm{Tr} (-1)^F x^{R-J} &= \int \frac{d\theta}{2\pi} \prod_{m=0}^\infty \frac{1-x^{\frac{4}{3}+2m}e^{i\theta}}{1-x^{\frac{2}{3}+2m}e^{-i\theta}} \cr
 &= 1-x^2-2 x^4-2 x^6-2 x^8+x^{12}+5 x^{14}+7 x^{16}+ \cdots \ . \label{indcs}
\end{align}
The integration over the $U(1)$ holonomy $\theta$ is equivalent to selecting gauge invariant operators.
From the invariance of the index, the expression \eqref{indcs} is also the index for the $\mathcal{N}=2$ Abelian Chern-Simons-matter theory in the non-relativistic limit with non-zero coupling constant.

Verma module of this theory is slightly more complicated because (at least in the $e \to 0$ limit) the scaling dimension takes a value of an integer and shows a multiplet shortening in the bosonic subalgebra as well. As discussed in section 3, there is a null vector at level 1, which is nothing but the Schr\"odinger equation (after Wick rotation):
\begin{align}
 -i\partial_t \Phi &= \frac{\Delta}{2m} \Phi \cr
-i\partial_t \Psi &= \frac{\Delta}{2m} \Psi \ .
\end{align}

\subsection{Non-relativistic limit of ABJM theory}
A new class of $\mathcal{N}=6$ superconformal field theories in $(1+2)$ dimensions which describes M2-brane on the orbifold has been proposed recently \cite{Aharony:2008ug}. The superconformal index in the relativistic case was studied in \cite{Bhattacharya:2008bja}. We take the non-relativistic limit of these theories and study their non-relativistic index.\footnote{In order to take the non-relativistic limit, we introduce the supersymmetric mass deformation \cite{Hosomichi:2008jd}\cite{Gomis:2008vc}.} As discussed in \cite{Bhattacharya:2008bja}, the coupling constant of the ABJM model is quantized and cannot be continuously deformed, but it is expected in the 't Hooft large $N$ limit, the index does not depend on the coupling constant and we can study the index structure in the zero-coupling constant limit.\footnote{In the zero coupling limit, the action is just given by the sum of Sch\"odinger action with the matter  representation specified below. The interacting action can be obtained by following the procedure introduced in \cite{Jackiw:1990mb}\cite{Leblanc:1992wu}. For our computation of the index, the explicit form of the interacting action is not necessary. See, however, the note added in section 6.}

We have presented the matter fields contributing to the (single particle) index in the zero coupling constant limit in Table \ref{tab:2}. The theory has $U(N) \times U(N)$ gauge symmetry and matter fields $\Phi_{12}^a$ (boson), $\Psi_{12}^a$ (fermion), $(a = 1,2)$ transform as fundamental representation in the first $U(N)$ group and anti-fundamental representation in the second gauge group. Similarly $\Phi_{21}^{\dot{a}},  \Psi_{21}^{\dot{a}}$  $(\dot{a}=1,2)$ transform as anti-fundamental representation in the first $U(N)$ group and fundamental representation in the second gauge group. The model possesses $SU(2) \times SU(2)$ global symmetry acting on $a$ and $\dot{a}$ independently.\footnote{These global symmetries will be eventually enhanced to $\mathcal{N}=6$ superconformal symmetry, but we will not go into further details.}

\begin{table}[tb]
\begin{center}
\begin{tabular}{c|c|c|c|c|c}
 Letters        & $U(N) \times U(N) $& $J$ & $R$ & $iD$ & $R-2J$ \\
 \hline
 $\Phi^a_{12}$&     $ N \times \bar{N} $         & $0$ & $2/3$ &$-1$ & 2/3 \\
  $\Phi^{\dot{a}}_{21}$&     $\bar{N} \times N $         & $0$ & $2/3$ &$-1$ & 2/3 \\
  $\Psi^{a*}_{21}$&     $\bar{N} \times N $         & $-1/2$ & $1/3$ & $-1$ & 4/3 \\ 
$\Psi^{\dot{a}*}_{12}$&     $N \times \bar{N} $         & $-1/2$ & $1/3$ & $-1$ & 4/3 \\ 
$P$&     $0 $         & $-1$ & $0$ & $-1$ & 2 \\ 
 \end{tabular}
\end{center}
\caption{List of the letters contributing to the index (hence $\Delta = 0$) for $\mathcal{N}=6$ ABJM Chern-Simons-matter theory. $a,\dot{a} = 1,2$ are indices for $SU(2) \times SU(2)$ global symmetry.}
\label{tab:2}
\end{table}%

The multi-particle index is obtained by using the effective matrix integration \cite{Kinney:2005ej}\cite{Nakayama:2006ur}\cite{Bhattacharya:2008bja}. It is given by the following matrix integration over the two $U(N)$ holonomy matrix $U_{a}$ as 
\begin{eqnarray}
I(x) = \int dU_1 dU_2 \exp\left(\sum_n \frac{1}{n} f_{12}(x^n) \mathrm{Tr}(U_1)^n \mathrm{Tr}(U_2^\dagger)^n + \frac{1}{n}f_{21}(x^n) \mathrm{Tr}(U_1^\dagger)^n \mathrm{Tr}(U_2)^n \right) \ ,
\end{eqnarray}
where $f_{12}$ and $f_{21}$ is the one-particle index for $\Phi_{12}^a$, $\Psi_{12}^{\dot{a}*}$ and  $\Phi_{21}^{\dot{a}}$, $\Psi_{21}^{{a}*}$ respectively.
\begin{align}
f_{12} = f_{21} = \frac{2x^{2/3}-2x^{4/3}}{1-x^2} \ .
\end{align}

The integration over the holonomy $U_1$ and $U_2$ is explicitly performable in the large $N$ limit, where the saddle point approximation is valid. The final result is
\begin{align}
I(x) &= \prod_{n=1}^\infty \frac{(1+x^{\frac{2}{3}n} + x^{\frac{4}{3}n})^2}{(1+x^{\frac{2}{3}n} - x^{\frac{4}{3}n})(1+x^{\frac{2}{3}n} + 3x^{\frac{4}{3}n})} \ . \cr & = 1+4 x^{4/3}-8 x^2+24 x^{8/3}-56 x^{10/3}+156 x^4-408 x^{14/3}+1076 x^{16/3} + \cdots \ . \label{inda}
\end{align}
The detail of the matrix integration may be found in the literatures \cite{Nakayama:2006ur}\cite{Bhattacharya:2008bja}.

To obtain more information of the theory, we utilize $SU(2) \times SU(2)$  global symmetry and further twist the superconformal index by introducing chemical potential $w^{2j^3_{(1)}} z^{2j^3_{(2)}}$. 
The index now becomes 
\begin{align}
&I(x,w,z) = \mathrm{Tr}(-1)^F x^{R-2J} w^{2j^3_{(1)}} z^{2j^3_{(2)}} \cr
&= \int dU_1 dU_2 \exp\left(\sum_{n=1}^{\infty} \frac{1}{n} f_{12}(x^n,w^n,z^n) \mathrm{Tr}(U_1)^n \mathrm{Tr}(U_2^\dagger)^n + \frac{1}{n}f_{21}(x^n,w^n,z^n) \mathrm{Tr}(U_1^\dagger)^n \mathrm{Tr}(U_2)^n \right) \ , \cr
\end{align}
where the single particle index is given by
\begin{align}
f_{12} &= \frac{x^{\frac{2}{3}}(w+w^{-1}) - x^{\frac{4}{3}}(z+z^{-1})}{1-x^2} \cr
f_{21} &= \frac{x^{\frac{2}{3}}(z+z^{-1}) -x^{\frac{4}{3}}(w+w^{-1}) }{1-x^2} \ .
\end{align}

Again the matrix integration can be performed explicitly in large $N$ limit as
\begin{align}
&I(x,w,z) = \prod_{n=1}^\infty \cr
&  \frac{(1-x^{2n})^2}{(1-x^{2n})^2 - \left(x^{\frac{2}{3}n}(w^n+w^{-n}) - x^{\frac{4}{3}n}(z^n+z^{-n})\right)\left(x^{\frac{2}{3}n}(z^n+z^{-n}) -x^{\frac{4}{3}n}(w^n+w^{-n})\right)} .
\end{align}
It would be interesting to construct maximal non-relativistic superconformal algebra of the ABJM model in the non-relativistic limit, and study the most general superconformal index with chemical potentials.

\section{Discussion}
In this paper, we have studied the highest-weight representation of the non-relativistic superconformal algebra and as an application we have defined the superconformal index. The index captures all the information which is invariant under exact marginal deformations. As a concrete example, we have computed the index for Chern-Simons-matter theories.

There are several possible future directions of this work. First of all, recently it has been shown that the non-relativistic superconformal algebra (especially $\mathrm{SSch}_2$) can be obtained from the reduction of relativistic superconformal algebra \cite{Sakaguchi:2008rx}\cite{Sakaguchi:2008ku} ($SU(2,2|1)$ corresponding to $\mathrm{SSch}_2$). It would be interesting to study the relation between the index defined for parent relativistic theories and the reduced non-relativistic theories. 

Another interesting direction would be related to the non-relativistic AdS-CFT correspondence recently advocated in the literatures \cite{Duval:1990hj}\cite{Son:2008ye}\cite{Balasubramanian:2008dm}\cite{Goldberger:2008vg}\cite{Barbon:2008bg}\cite{Wen:2008hi}\cite{Herzog:2008wg}\cite{Maldacena:2008wh}\cite{Adams:2008wt}. As we mentioned in the introduction, the comparison of the index would be the first check of the proposed duality. Since the gravity dual for the relativistic Chern-Simons-matter theory studied in section 5.2 is known \cite{Aharony:2008ug}, it would be interesting to investigate its non-relativistic limit realized in supergravity/string theory. Some supergravity backgrounds corresponding to non-relativistic conformal field theories have been studied in \cite{Herzog:2008wg}\cite{Maldacena:2008wh}\cite{Adams:2008wt}.

\

\

{\bf Note added}

Recently, it has been shown that the naive non-relativistic limit of the ABJM theory discussed in section 5 is not invariant under the dynamical superysmmetry $Q_2$\cite{Nakayama:2009cz}\cite{Nakayama:2009ku}. However, we can deform the fermionic potential so that the dynamical supersymmetry is preserved, so we regard the computation done in section 5 as an index for such a deformed theory. It is also possible to study the index for the non-relativistic ABJM theory with $14$ supercharges studied in \cite{Nakayama:2009cz}. It turns out that the computation of the index is the same as that in section 5 by replacing $\Phi^{\dot{a}}_{21}$ with $\Phi^{\dot{a}*}_{21}$ and $\Psi^{\dot{a}*}_{12}$ with $\Psi^{\dot{a}}_{12}$. The index itself is still given by \eqref{inda}.

\section*{Acknowledgements}
The research of Y.~N. is supported in part by NSF grant PHY-0555662 and the UC Berkeley Center for Theoretical Physics.

\end{document}